\documentclass[prl,showpacs,twocolumn,superscriptaddress]{revtex4}
\usepackage{graphicx}

\begin{document}

\title{Abrupt Change in Radiation-Width Distribution for $^{147}$Sm Neutron
Resonances}
\author{P.~E.~Koehler}
\thanks{corresponding author}
\affiliation{Physics Division, Oak Ridge National Laboratory, Oak Ridge, TN 37831, USA}
\author{R. Reifarth}
\affiliation{Goethe Universit\"{a}t, Frankfurt Am Main, Germany}
\author{J. L. Ullmann}
\affiliation{Los Alamos National Laboratory, Los Alamos, NM 87454, USA}
\author{T. A. Bredeweg}
\affiliation{Los Alamos National Laboratory, Los Alamos, NM 87454, USA}
\author{J. M. O'Donnell}
\affiliation{Los Alamos National Laboratory, Los Alamos, NM 87454, USA}
\author{R. S. Rundberg}
\affiliation{Los Alamos National Laboratory, Los Alamos, NM 87454, USA}
\author{D. J. Vieira}
\affiliation{Los Alamos National Laboratory, Los Alamos, NM 87454, USA}
\author{J. M. Wouters}
\affiliation{Los Alamos National Laboratory, Los Alamos, NM 87454, USA}
\date{\today }

\begin{abstract}
We obtained total radiation widths of \textit{s}-wave resonances through $%
\mathcal{R}$-matrix analysis of $^{147}$Sm($n,\gamma $) cross-sections.
Distributions of these widths differ markedly for resonances below and above 
$E_{n}=300$ eV, in stark contrast to long-established theory. We show that
this change, as well as a similar change in the neutron-width distribution
reported previously, are reflected in abrupt increases in both the average $%
^{147}$Sm($n,\gamma $) cross section and fluctuations about the average near
300 eV. Such effects could have important consequences for applications such
as nuclear astrophysics and nuclear criticality safety.
\end{abstract}

\maketitle

In this letter, we show that total-radiation-widths ($\Gamma _{\gamma }$)
extracted from $\mathcal{R}$-matrix analysis of $^{147}$Sm($n,\gamma $)
cross sections reveal an abrupt change in the shape and average value of the 
$\Gamma _{\gamma }$ distribution near $E_{n}=300$ eV. These observations are
in stark contrast with theoretical expectations that both quantities should
remain essentially constant across the resonance range and beyond.

The effect reported herein occurs very near the same energy as previously
reported abrupt changes in the $\alpha $-particle strength-function ratio 
\cite{Ko2004} for the two \textit{s}-wave spin states and the shape of the
reduced-neutron-width ($\Gamma _{n}^{0}$) distribution \cite{Ko2007}. Due to
the difficulty of measuring the very small $\alpha $ widths, the former
effect was of limited statistical significance. However, the effect in the $%
\Gamma _{n}^{0}$ data was established at about the 99\% confidence level
using several different tests. These two previous effects remain unexplained.

As we will show below, changes in the $\Gamma _{\gamma }$ distribution shape
and average are established with very high confidence. That three such
deviations from theoretical expectations could occur by chance in the same
nuclide at the same energy must be vanishingly small. Therefore, it is
virtually certain that significant departure from standard theory has been
observed and that all three effects may have a common origin. Given the
relative paucity of high-quality $\Gamma _{\gamma }$ data, similar effects
may exist for other nuclides and, if so, could have far-reaching
consequences for both basic and applied nuclear physics. For example, as we
show below, these changes in the $\Gamma _{\gamma }$ and $\Gamma _{n}^{0}$
distributions are reflected in abrupt increases in both the average $^{147}$%
Sm($n,\gamma $) cross section and fluctuations about the average that cannot
be explained by the nuclear statistical model. As this theory is used to
calculate cross sections for applications, similar differences in other
nuclides could have important impacts in nuclear astrophysics and nuclear
criticality safety.

It is expected that $\Gamma _{\gamma }$ distributions for medium to heavy
nuclides should be very narrow, and essentially constant across the
resonance energy region. Expectation \cite{Po56} that the $\Gamma _{\gamma }$
distribution should be very narrow arises from i) the very complex wave
functions of states at high excitation characteristic\ of neutron thresholds
and ii) the very large number of channels for $\gamma $ decay. Condition i)
results in partial $\gamma $-decay widths $\Gamma _{i\gamma }$ for each
channel $i$ following a Porter-Thomas distribution (PTD) \cite{Po56} (a $%
\chi ^{2}$ distribution with one degree of freedom, $\nu _{i\gamma }=1$).
Condition ii) results in the expectation that total $\gamma $-decay widths $%
\Gamma _{\gamma }=\sum_{i=1}^{n}\Gamma _{i\gamma }$ will follow a $\chi ^{2}$
distribution with degrees of freedom given by the number of
independently-contributing channels $n\equiv \nu _{\gamma
}=\sum_{i=1}^{n}\nu _{i\gamma }$. As $n\sim 100$, $\Gamma _{\gamma }$
distributions can indeed be very narrow.

That the $\Gamma _{\gamma }$ distribution should remain fairly constant
arises from consideration of the physics of radiative transitions in nuclei
as implemented in the nuclear statistical model, which should apply for
nuclides in this mass range at these excitation energies. Partial radiation
widths $\Gamma _{i\gamma }$ for transitions from resonances $i$ to
individual final levels are characterized by average values

\begin{equation}
\langle \Gamma _{i\gamma }\rangle =\frac{f_{XL}(E_{\gamma })E_{\gamma
}^{(2L+1)}}{\rho (E_{i},J_{i},\pi _{i})},  \label{GgExpectationValueEquation}
\end{equation}%
where $E_{\gamma }$ is the $\gamma $-ray energy, $\rho (E_{i},J_{i},\pi
_{i}) $ is the density of resonances with spin $J_{i}$ and parity $\pi _{i}$
at energy $E_{i}$, and $f_{XL}(E_{\gamma })$ is the photon strength function
(PSF) for transitions of type $X$ (electric or magnetic) and multipolarity $%
L $. As the resonance-energy range is rather small and all quantities are
smooth functions of energy, no abrupt changes are expected.

Resonance $\Gamma _{\gamma }$ values typically have been determined only for
a relatively small subset of observed resonances, and quite often have
fairly large uncertainties. On the other hand, $\Gamma _{\gamma }$ data may,
in principle, be a more sensitive tool for testing theory than $\Gamma
_{n}^{0}$ data, for at least two reasons. First, because $\Gamma _{\gamma }$
distributions are much narrower than $\Gamma _{n}^{0}$ (which are predicted
to follow the PTD, $\nu _{n}=1$), it is much easier to detect a change in
distribution shape in the former case with a limited amount of data. Second,
systematic effects due to missed resonances should be negligible, or at
least much smaller, for $\Gamma _{\gamma }$ compared to $\Gamma _{n}^{0}$
data. It is a well-known fact that all experiments miss some resonances
having small neutron widths, and it is well established that neglecting this
effect can cause significant systematic errors in discerning the shape of
the $\Gamma _{n}^{0}$ distribution from the data. Because the range of $%
\Gamma _{\gamma }$ values is much smaller, and because they are, in general,
uncorrelated with $\Gamma _{n}^{0}$ (we have verified this for the data used
herein), missed resonances should have random $\Gamma _{\gamma }$ values,
and hence no correction for missed resonances is needed while determining
the shape of the $\Gamma _{\gamma }$ distribution from the data.

All resonances should have the same parity to perform a valid test of the
theory. $^{147}$Sm is ideal in this regard because it is at the peak of the 
\textit{s}- and minimum of the \textit{p}-wave neutron strength functions,
so all observed resonances at the low energies used in our analysis should
be \textit{s} wave. The probability of a \textit{p}-wave resonance being
included in our analyses can be estimated from the average resonance
parameters in Ref. \cite{Mu2006} (\textit{p}-wave neutron strength function $%
10^{4}S_{1}=0.9$ and \textit{s}-wave average resonance spacing $D_{0}=5.7$
eV), the usual assumption that the \textit{p}-wave average resonance spacing 
$D_{1}=$ $\frac{1}{3}D_{0}$, and the threshold used in our analysis (see
below). From these values, it is easy to show that the probability of a 
\textit{p}-wave resonance being included in our analyses is extremely low,
being less than $2\times 10^{-7}$.

Extracting $\Gamma _{\gamma }$ values from measured cross sections also
requires knowing the resonance spins. For $^{147}$Sm ($I^{\pi }=7/2^{-}$),
two \textit{s}-wave spins are possible, $J^{\pi }=3^{-}$ and $4^{-}$. We
overcame this potential problem by making the measurements with the Detector
for Advanced Neutron Capture Experiments (DANCE) \cite{He2001}, with which
we were able to determine firm spin assignments for all resonances used
herein.

Details of the experiment have been reported elsewhere \cite{Ko2007}. DANCE
is a $4\pi $ $\gamma $-ray detector comprised of 160 BaF$_{2}$
scintillators, each coupled to its own photomultiplier tube, the outputs of
which were inputted to two transient digitizers each. In this way, waveforms
for each detector were recorded for each neutron beam pulse and analyzed in
real time to detect peaks, whose shape and time stamp were written to disk.
A 10.4-mg metallic sample, enriched to 97.93\% in $^{147}$Sm and mounted on
a thin Al backing, was placed in the center of DANCE. A well collimated
neutron beam from a water moderator at the Manuel Lujan, Jr. Neutron
Scattering Center at the Los Alamos Neutron Science Center was incident on
this target.

Neutron energies were determined using the time-of-flight technique during
replay of the data. Cuts were applied to reduce backgrounds and restrict
events to those in the range expected from $^{147}$Sm($n,\gamma $)
reactions. Separate measurements were made with a blank Al backing foil.
Neutron flux was redundantly measured using three different sample/detector
combinations. Flux-normalized sample-out counts were subtracted from the
sample-in data. Resulting neutron-capture cross sections $\sigma _{\gamma
}(E_{n})$ in the unresolved region are in agreement with the most recent
high-accuracy data \cite{Wi93} to within the uncertainties.

\begin{figure}[b]
\includegraphics[clip,width=1.0\columnwidth]{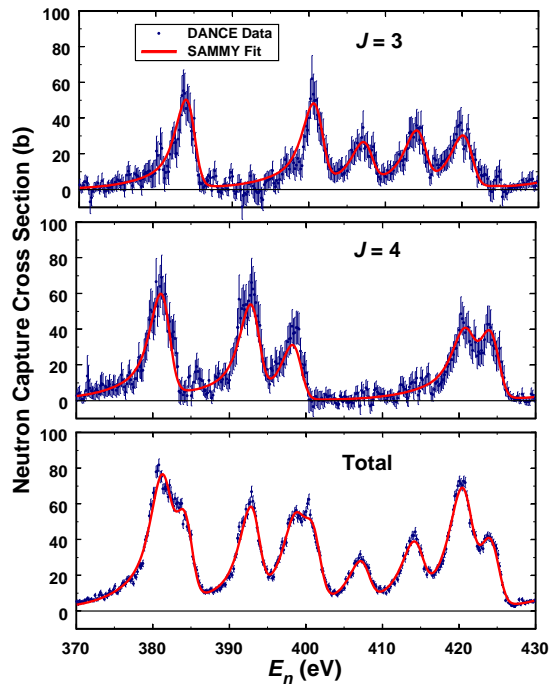} 
\vspace*{-0.3cm}
\caption{(Color online) A small part of our $^{147}$Sm($n,\protect\gamma $)
cross sections versus neutron energy. Top, middle, and bottom panels show $%
J=3$, 4, and total capture cross sections, respectively. See text for
details.}
\label{SpinSepSigs}
\end{figure}

As explained in Ref. \cite{Ko2007}, $\gamma $-ray multiplicity (the number
of $\gamma $-rays emitted following neutron capture) information measured
with DANCE makes this detector an excellent resonance "spin meter". The
technique invented in Ref. \cite{Ko2007} was further developed in Ref. \cite%
{Be2011}. We used the least-squares version in the latter reference to
determine spin-separated yields as functions of neutron energy $q^{J}(E_{n})$
for the two \textit{s}-wave spins. Spins of all resonances analyzed herein
could be determined from these yields by inspection. We also used these
yields to calculate spin-separated neutron-capture cross sections $\sigma
_{\gamma }^{J}(E_{n})$, e.g., $\sigma _{\gamma }^{3}(E_{n})=\sigma _{\gamma
}(E_{n})q^{3}(E_{n})/(q^{3}(E_{n})+q^{4}(E_{n}))$. Example cross sections
are shown in Fig. \ref{SpinSepSigs}. The spin-separated cross sections were
crucial for obtaining $\Gamma _{\gamma }$ values for resonances which were
not fully resolved in the DANCE data.\bigskip 

The $\mathcal{R}$-matrix code SAMMY \cite{La2008} was used to fit the $%
^{147} $Sm($n,\gamma $) cross-section data and obtain resonance parameters.
Because natural widths ($\Gamma _{n}+\Gamma _{\gamma }$) of the resonances
were almost always smaller than the sum of the experiment resolution plus
Doppler broadening, fitting the capture data alone yields capture kernels,

\begin{equation}
K_{\gamma }=g_{J}\Gamma _{n}\Gamma _{\gamma }/(\Gamma _{n}+\Gamma _{\gamma
}),  \label{CaptureKernelEquation}
\end{equation}%
(where $g_{J}=(2J+1)/[(2I+1)(2j+1)]$ is the statistical factor for
resonance, target, and neutron spins $J$, $I$, and $j$, respectively) and
not the individual widths. However, neutron widths are available from
resonance analysis of total cross section data. Therefore, $\Gamma _{\gamma
} $ values were determined by fitting our data using the $g_{J}\Gamma _{n}$
values from Ref. \cite{Mi81}.

There are two components to the total uncertainty $\Delta \Gamma _{\gamma }$
for each $\Gamma _{\gamma }$ value. First, there is the contribution due to
fitting the data, which was calculated by SAMMY. Second, there is a
contribution due to the uncertainty in the neutron width, which was derived
in the standard manner using Eq. \ref{CaptureKernelEquation} and the $\Delta
\Gamma _{n}$ values from Ref. \cite{Mi81}. These two components were added
in quadrature to obtain total uncertainties. As can be seen from Eq. \ref%
{CaptureKernelEquation}, if $\Gamma _{n}$ is significantly larger than $%
\Gamma _{\gamma }$, then the capture kernel is relatively insensitive to $%
\Gamma _{n}$ and the resulting $\Delta \Gamma _{\gamma }$ is essentially
that calculated by SAMMY. However, as $\Gamma _{n}$ decreases, $\Delta
\Gamma _{\gamma }$ increases. In the limit $\Gamma _{n}\ll \Gamma _{\gamma }$%
, the capture kernel is essentially equal to $g_{J}\Gamma _{n}$, and hence $%
\Gamma _{\gamma }$ cannot be determined. Therefore, some limit has to be
imposed on the subsequent analysis. Experience has shown $\Gamma _{n}\geq
\Gamma _{\gamma }/2$ to be reasonable, and hence we limited the subsequent
analyses to such resonances. We have repeated the analyses described below
using other reasonable limits (e.g., $\Delta \Gamma _{\gamma }/\Gamma
_{\gamma }<10\%$, 5\%) and obtained essentially the same results.

\begin{figure}[b]
\includegraphics[clip,width=1.0\columnwidth]{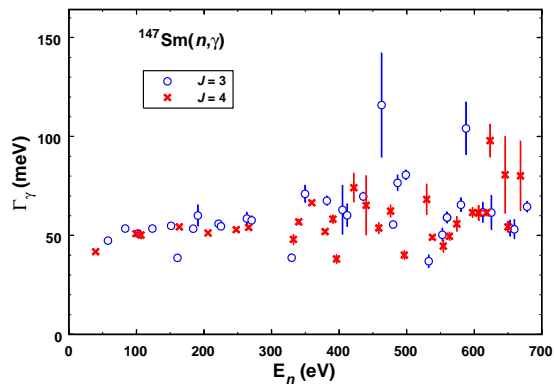} \vspace*{-0.3cm%
}
\caption{(Color online) Resonance $\Gamma _{\protect\gamma }$ values versus
energy from the SAMMY fits to our data for the 62 resonances meeting the
criteria discussed in the text. Values for $3^{-}$ and $4^{-}$ resonances
are shown as blue cirlces and red X's, respectively.}
\label{GgVsE}
\end{figure}

The $\Gamma _{\gamma }$ values for the 62 (out of 112 observed) resonances
below 700 eV meeting the above criteria are shown as a function of resonance
energy in Fig. \ref{GgVsE}. As can be seen in this figure, there is no
discernible difference in $\Gamma _{\gamma }$ values for the two \textit{s}%
-wave spins, and the $\Gamma _{\gamma }$ distribution becomes noticably
broader for $E_{n}>$ 300 eV. Therefore, we combined data for the two spins
in subsequent analyses, and divided the data into two groups; $E_{n}<300$ eV
and $300<E_{n}<700$ eV.

\begin{figure}[b]
\includegraphics[clip,width=1.0\columnwidth]{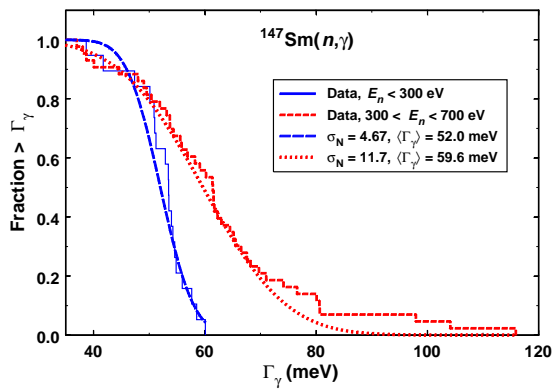} \vspace*{%
-0.3cm}
\caption{(Color online) Cumulative $\Gamma _{\protect\gamma }$
distributions. Shown are the fraction of resonances with $\Gamma _{\protect%
\gamma }$ larger than a given value versus the value. Staircase plots depict
the measured data whereas smooth curves show Gaussian distributions from the
ML analyses.}
\label{GgCumDists2ERegs}
\end{figure}

Cumulative $\Gamma _{\gamma }$ distributions for the two energy regions are
shown in Fig. \ref{GgCumDists2ERegs}. We performed several tests \cite{Co80}
to discern the statistical significance of the change in distribution shape
which is evident in this figure.

The median (variance) test indicates the null hypothesis that medians
(variances) of the two distributions are the same can be rejected at the
99.8\% (99.9\%) confidence level. Similarly, the Smirnov and Cramer-von
Mises two-sample tests reveal the null hypothesis that data in the two
energy regions were sampled from the same population can be rejected with 
$%
>99\%$ and $>99.9\%$ confidence, respectively. In
essence, all these statistical tests indicate that the change in the $\Gamma
_{\gamma }$ distribution evident in Fig. \ref{GgCumDists2ERegs} is highly
statistically significant.

Theoretical interpretation of this change may be aided by estimation of
distribution parameters for the two regions. To this end, we used the
maximum likelihood (ML) method. As noted above, $\Gamma _{\gamma }$ data are
expected to follow a $\chi ^{2}$ distribution with many degrees of freedom, $%
\nu _{\gamma }\sim 100$. For such large values of $\nu _{\gamma }$, a $\chi
^{2}$ distribution is very close to Gaussian in shape. One advantage of
using a Gaussian rather than $\chi ^{2}$ distribution for the analysis is
that uncertainties $\Delta \Gamma _{\gamma }$ can easily be included \cite%
{As66}.

Therefore, we used the technique described in Ref. \cite{As66} to estimate
most likely values for the means $\langle \Gamma _{\gamma }\rangle $ and
standard deviations $\sigma _{N}$ of the $\Gamma _{\gamma }$ distributions
in the two energy regions. Resulting ML estimates are $\sigma _{N}=4.67\pm
0.81$, $\langle \Gamma _{\gamma }\rangle =52.0\pm 1.1$, and $\sigma
_{N}=11.7\pm 1.5$, $\langle \Gamma _{\gamma }\rangle =59.6\pm 2.0$, for the
lower- and upper-energy regions, respectively. Hence, the ML results also
indicate that $\Gamma _{\gamma }$ distributions in the two energy regions
are significantly different. Translated to $\chi ^{2}$ distributions, these
ML results lead to $\nu _{\gamma }=241$ and $51$ for the $\Gamma _{\gamma }$
distributions in the lower and upper energy regions, respectively.

Because our dividing energy is slightly different than that used in Ref. 
\cite{Ko2007}, our new $\Gamma _{n}^{0}$ values for smaller resonances are
slightly different, and the ML technique of Ref. \cite{Ko2010a} is better
than that used in Ref. \cite{Ko2007}, we reanalyzed the $\Gamma _{n}^{0}$
data using the technique of Ref. \cite{Ko2010a} to obtain $\nu _{n}$ values
for the two energy regions. As explained in the Ref. \cite{Ko2010a}, the ML
analysis technique employs an energy dependent threshold to properly account
for the effect of missed small resonances.  The results given in Table \ref%
{AveResParTable} were obtained with a threshold $g_{J}\Gamma _{n}^{0}$ $\geq
3.3\times 10^{-4}E_{n}$, where $g_{J}\Gamma _{n}^{0}$ is given in meV for $%
E_{n}$ in eV. As explained in the Ref. \cite{Ko2010a}, this same threshold
excludes \textit{p}-wave resonances from the analysis with equal
effectiveness at all energies. We then applied the technique of Ref. \cite%
{Fu65} (using the same threshold, and modified to work for any $\nu _{n}$)
to obtain average \textit{s}-wave resonance spacings $D_{0}$ and neutron
strength functions $S_{0}$, corrected for missed small resonances. These
values, along with parameters for the $\Gamma _{\gamma }$ distributions, are
given in Table \ref{AveResParTable}. Our resulting $\nu _{n}$ values are
consistent with those of Ref. \cite{Ko2007} and so confirm that the $\Gamma
_{n}^{0}$ distribution also changes shape near 300 eV.

\begin{table*}[tbp] \centering%
\caption{Average parameters for $^{147}$Sm+\textit{n}
resonances.\label{AveResParTable}}%
\begin{tabular}{cccccc}
\hline\hline
Region (eV) & $\langle \Gamma _{\gamma }\rangle $ (meV) & $\nu _{\gamma }$ & 
$10^{4}S_{0}$ & $D_{0}$ (eV) & $\nu _{n}$ \\ \hline
0 - 300 & $52.0\pm 1.1$ & $241\pm 42$ & $4.56\pm 0.94$ & $5.78\pm 0.44$ & $%
1.04_{-0.31}^{+0.33}$ \\ 
300 - 700 & $59.6\pm 2.0$ & $51.0\pm 6.5$ & $4.53\pm 0.81$ & $6.18\pm 0.40$
& $2.67_{-0.57}^{+0.60}$ \\ \hline\hline
\end{tabular}%
\end{table*}%

\begin{figure}[b]
\includegraphics[clip,width=1.0\columnwidth]{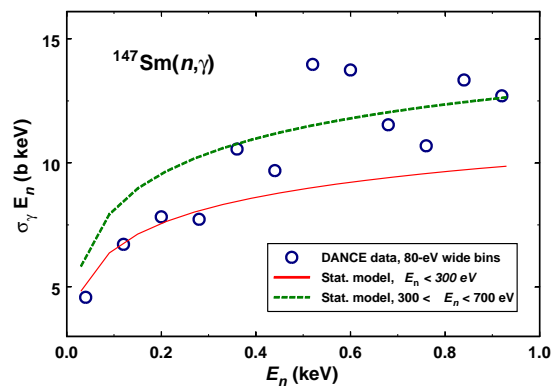} \vspace*{%
-0.3cm}
\caption{(Color online) Open blue circles depict our DANCE $^{147}$Sm($n,%
\protect\gamma $) cross sections averaged over 80-eV-wide bins. Error bars
corresponding to one-standard-deviation statistical uncertainties are
smaller than the symbols. The solid red and dashed green curves show results
of statisical model calculations based on the average resonance parameters
for the lower- and upper-energy regions, respectively. See text for details.}
\label{AveCSvE}
\end{figure}

That these changes in the $\Gamma _{\gamma }$ and and $\Gamma _{n}^{0}$
distributions are mirrored in the $^{147}$Sm($n,\gamma $) cross section is
shown in Fig. \ref{AveCSvE}, in which our DANCE data averaged over
80-eV-wide bins are shown. The bin width must be wide enough to contain
several resonances so that the large fluctuations in resonance sizes are
damped, but not so large that the change near 300 eV is averaged out. As the
average resonance spacing is about 6 eV, the chosen bins should contain
about 13 resonances on average, which should be a good compromise. For
comparison, the typical rule of thumb for statistical model calculations is
that the energy interval contain at least 10 resonances.

Also shown in Fig. \ref{AveCSvE} are two statistical model calculations
based on the average resonance parameters given in Table \ref{AveResParTable}
for the two energy regions. As all statistical model codes of which we are
aware assume the PTD for $\Gamma _{n}^{0}$, and essentially a single,
constant $\langle \Gamma _{\gamma }\rangle $, we wrote our own simple code
which randomly samples over $\chi ^{2}$ distributions with parameters given
in Table \ref{AveResParTable}. As can be seen in Fig. \ref{AveCSvE}, there
is a substantial, fairly abrupt, change in the measured cross section near
300 eV, and calculations based on the parameters for each region are in good
agreement with the data in that region, but inconsistent with data in the
other region. It also is evident that fluctuations in the data about the
theoretical value are larger in the upper-energy region, which is consistent
with the interpretation \cite{Ko2004,Ko2007}\ of a non-statistical effect in
this region. The cross section in the upper-energy region is approximately
30\% larger than calculated using parameters for the lower-energy region.
Our calculations indicate that about one third of this increase is due to
the changes in the $\Gamma _{\gamma }$ distribution, and the remaining two
thirds is mainly due to width-fluctuation effects in the neutron channel.

To our knowledge, there is no model which can explain the two previously
reported or the current effects in $^{147}$Sm+$n$ widths near 300 eV. The
above $\Gamma _{\gamma }$ results could be interpreted as a decrease in the
number of effectively independent channels by 190, or a decrease in the
degrees of freedom for each channel by approximately a factor of 1/5,
between the two energy regions. Given the extremely limited $\Gamma _{\alpha
}$ data available, the paucity of high-quality $\Gamma _{\gamma }$ data, and
the near universal practice of assuming $\Gamma _{n}^{0}$ follow the PTD,
similar effects could exist in other nuclides. In addition to interest in
understanding the underlying theory, such effects may be important to, for
example, nuclear astrophysics and nuclear criticality safety, in which
models often are used to calculate important quantities beyond the reach of
measurement. Because our results do not agree with predictions and
assumptions of these models, it is prudent to assume that quantities
predicted by these models may be more uncertain than previously thought.
Similar quality data on other nuclides likely will be needed before the
origin and extent of the effects presented herein can be understood.

\begin{acknowledgments}
The authors would like to thank R. R. Winters for useful discussions. This
work was supported by the Office of Nuclear Physics of the U.S. Department
of Energy under Contract No. DE-AC05-00OR22725 with UT-Battelle, LLC. This
work has benefited from the use of the LANSCE facility at Los Alamos
National Laboratory which was funded by the U.S. Department of Energy and
operated by the University of California under Contract W-7405-ENG-36.
\end{acknowledgments}

\bibliographystyle{prsty}
\bibliography{ACOMPAT,pauls}

\newif\ifabfull\abfulltrue

\end{document}